\documentclass[preprint, 5p,times,twocolumn]{elsarticle}

\usepackage{xcolor}
\definecolor{neonblue}{RGB}{137, 207, 240}
 \usepackage{xurl}
 \usepackage{hyperref}
\usepackage{hyperref}
\hypersetup{
  colorlinks=true,
  linkcolor=neonblue, % Neon blue color
  citecolor=neonblue,
  urlcolor=neonblue,
}

\usepackage{amssymb}
\usepackage{amsmath}
\usepackage{bm}
\usepackage{multirow}
\usepackage{booktabs}

\journal{Aerospace Science and Technology (AESCTE)}

\begin{document}

\begin{frontmatter}


\title{Thrust vector control and state estimation architecture for low-cost small-scale launchers}
%\title{Integrated architeture for state estimation and thrust vector control in low-cost small-scale launchers}


 \author[IDMEC]{Pedro dos Santos\corref{cor}}
 \ead{pedrodossantos31@tecnico.ulisboa.pt}
 \cortext[cor]{Corresponding author}

\author[IDMEC]{Paulo Oliveira} 
%\ead{paulo.j.oliveira@tecnico.ulisboa.pt}
 
 %\affiliation[IST]{organization={Instituto Superior Técnico, Universidade de Lisboa},
%             addressline={Av. Rovisco Pais 1},
             %city={},
%             postcode={1049-001},
%             state={Lisbon},
%           country={Portugal}}
           
  \affiliation[IDMEC]{organization={IDMEC, Instituto Superior Técnico, Universidade de Lisboa},
             addressline={Av. Rovisco Pais 1},
             %city={},
             postcode={1049-001},
             state={Lisbon},
             country={Portugal}             
           }
\begin{abstract}
This paper proposes an integrated architecture for Thrust Vector Control (TVC) and state estimation for low-cost small-scale launchers, naturally unstable, and propelled by a solid motor. The architecture is based on a non-linear, six-degrees-of-freedom model for the generic thrust-vector-controlled launcher dynamics and kinematics, deduced and implemented in a realistic simulation environment. For estimation and control design purposes, a linearized version of the model is proposed. Single-nozzle TVC actuation is adopted, allowing for pitch and yaw control, with the control law being derived from the Linear Quadratic Regulator (LQR) with additional integral action (LQI). The control system is implemented through gain scheduling. Full state estimation is performed  resorting to complementary kinematic filters, closely related to linear Kalman filtering theory. The architecture, composed by the navigation and control systems, is tested in simulation environment, demonstrating satisfactory attitude tracking performance and robustness to both external disturbances and model uncertainties. 

\end{abstract}




\end{frontmatter}


\section{Introduction}
\label{sec:intro}

With the increasing number of small satellite manufacturers, namely of the ``cubesat" cathegory, the market for cost-efficient small-scale launchers tends to grow larger \cite{orbitaltoday}. Traditionally, these satellites were launched as secondary payload in large-scale launchers, where the mission profile would be tailored given the primary payload client's requirements. Although the ``rideshare'' scenario allows for cost reduction, dedicated small-scale launchers give their small payload clients more flexibility, for instance by providing the opportunity to select the desired final orbit and launch date \cite{spacetecpartners_2022}.

In addition to orbital class small launchers, sub-orbital launchers have long been used to conduct scientific experiments and take measurements in high altitudes (from tenths to hundreds of kilometers) and in microgravity conditions,  hence their common denomination of ``sounding rockets'' \cite{nasa_sr}. More recently, sub-orbital transportation and space tourism motivated a market increase which impacts the overall need for cost-effective, dedicated sub-orbital launchers, with an emphasis on reusability \cite{spacetourism}.

To meet specific mission requirements, in terms of stability and trajectory, launch vehicles must have a dedicated Guidance, Navigation, and Control (GN\&C) system. This system is responsible for determining on board the trajectory to be followed and commanding the required attitude (or orientation) over time (Guidance), for estimating the state vector, composed by position, velocity, and attitude (Navigation), and for calculating the necessary actuation inputs to achieve the desired attitude (Control). In this paper, the focus is on navigation and control, with an integrated architecture for attitude control and state estimation, suitable for low-cost small-scale launchers, being proposed.  As for the actuation method, Thrust Vector Control (TVC), or thrust vectoring for short, is selected. 

TVC is used by most launch vehicles and works by redirecting the thrust vector in order to create a control torque \cite{nasacontroldesign}. When TVC is achieved through a single gimballed nozzle, which is a suitable configuration for small-scale launchers, it can only impact the pitch and yaw angles, whereas roll has to be controlled by an additional system, if needed.  With respect to other actuation techniques, such as actively controlled fins, TVC allows for a wider range of operating conditions and provides better efficiency \cite{prop_sutton}. 

Solid motors are the most common propulsion technology in small-scale launchers due to their Intercontinental Cruise Ballistic Missile (ICBM) heritage \cite{Kulu2021} and associated low production costs, which enables rapid and responsive launch missions \cite{zhang}. Therefore, it has been selected as reference for this work. Contrarily to liquid or hybrid engines, which seem to be the future trend in small launchers \cite{Kulu2021}, solid motors do not possess throttle capability. This means that thrust cannot be controlled and, consequently, control authority is reduced.

The control system design tends to be very conservative in the aerospace industry \citep{Tewari}. Restricting the dynamic analysis to accommodate more sophisticated control design techniques risks the later realization that such restrictions would have to be lifted and would invalidate the control design \citep{ref_prof}. Due to the highly non-linear dynamics and to the time-varying nature of the parameters, such as aerodynamic and inertial, the applicability of linear control techniques relies on the linearization of the system at several operating points. The design is then focused in each linear model and the resulting controller gains are changed during the flight through a technique called gain scheduling, as in \cite{bei}. 

Classic and linear control solutions, based on thrust vectoring, can be found in \cite{nasacontroldesign, nasathird, f9}. These include Proportional-Integrative-Derivative (PID) control and pole placement techniques, both with time-varying gains. Although widely used, PID control has its downsides when it comes to model uncertainty robustness and external disturbances rejection.

Still in the linear domain, the use of optimal controllers, such as the Linear Quadratic Regulator (LQR), provides some degree of robustness and ensures a (sub-)optimal trajectory tracking solution for a given cost function. In \cite{lqrrocket}, LQR is used to address the attitude control problem, and in \cite{kisabo} an LQG algorithm is proposed for state estimation and control, with both restricting the analysis and design to the pitch plane at a single operating point. 
 
Non-linear techniques have also been proposed for launch vehicle control and estimation \cite{mooji, celani}, and come with the advantage of ensuring a global solution, not dependant on the specific mission nor vehicle. However, these methods all have particular design characteristics which hinder the application of standardized, well-established, verification and validation procedures \cite{nasacontroldesign, ref_prof, nasathird}.

Although several solutions to the launcher control problem can be found in the literature, many fail to capture all the relevant dynamics and/or oversimplify the problem, while most assume full-state knowledge, creating a considerable gap between theoretical design and implementation. Hence, the main contribution of this paper is a robust architecture, which integrates both the navigation and control systems, that can be easily reproduced and implemented in low-cost launchers and relies on readily available components. 

For the navigation system, complementary kinematic filters, relying on Kalman filtering theory, are proposed to fuse the sensor readings and obtain filtered, unbiased, full-state estimates. The use of complementary kinematic filters allows to obtain a solution which does not require extensive tuning for each specific mission and to use linear Kalman filtering, avoiding excessive computation effort.

As for the control system, LQR control is proposed with additional integral action (LQI) to increase robustness and provide a null attitude tracking error. The gains are obtained for different operating points of the reference trajectory, to be scheduled during flight with an altitude-based linear interpolation.

This paper is organized as follows: the problem is presented in Section \ref{sec:prob}. Some notation is detailed in Section \ref{sec:notation}. The physical model is shown in Section \ref{sec:model}. The linear state-space representation is derived in Section \ref{sec:linearss}. The proposed architecture is explained in Section \ref{sec:arch}. The navigation and control systems are detailed in Sections \ref{sec:navigation} and \ref{sec:control}, respectively. Section \ref{sec:implement} shows the implementation in simulation of the architecture, as well as the reference vehicle and mission used for validation. In Section \ref{sec:linanalysis}, a linear domain analysis of the system follows, and in Section \ref{sec:simres} the simulation results are presented and discussed. Finally, in Section \ref{sec:conclusions}, final remarks and conclusions are drawn.

\section{Problem statement}
\label{sec:prob}

This paper presents an integrated architecture for the attitude control and state estimation of thrust-vector-controlled, small-scale launch vehicles without aerodynamic fins. In the absence of fins, launch vehicles are naturally unstable since the centre of mass is located aft of the centre of pressure \citep{cornelisse}. Hence, the need for a control system is evident. Besides stabilizing the plant, this system is also responsible for rejecting external disturbances, such as wind gusts, and actively correcting the trajectory. To implement a control system, it is necessary to have accurate estimates on the state vector of the vehicle, imposing the need for a navigation system composed by sensors and estimators.

As a single gimballed nozzle is assumed for actuation, the spinning motion of the vehicle cannot be controlled via thrust vectoring. In this way, the architecture has to be designed to provide pitch and yaw control in the presence of spinning motion, making an additional roll control system only necessary to limit the spin velocity to an admissible value.

Finally, since solid propulsion technology is assumed, thrust is not controllable. This means that the control authority is reduced and that the implementation of the architecture will depend on the thrust curve for the specific vehicle and mission.

\section{Notation}
\label{sec:notation}
This paper uses bold lowercase and bold uppercase symbols to represent vector and matrices, respectively, superscript $T$ to denote the transpose, superscript $-1$ to denote the inverse, $\mathbf{I_{n}}$ to represent the identity matrix of dimension $n$, and $\mathbf{0_{mxn}}$ to represent the null matrix of dimension $m$ by $n$.

\section{Physical model}
\label{sec:model}

In this section, the dynamics and kinematics of a generic launch vehicle with a single gimballed nozzle are provided. To derive the physical model some assumptions are used: the launch vehicle is assumed to be a rigid body; it is assumed to be axially symmetric, as well as the mass allocation; and the flat Earth model is used, neglecting Earth's curvature and rotation. All these assumptions are considered valid for first stage design of the architecture and don't compromise its overall structure when reproducing it in a real case scenario.

\subsection{Reference frames}
\begin{figure}[!htpb]
    \centering
    \includegraphics{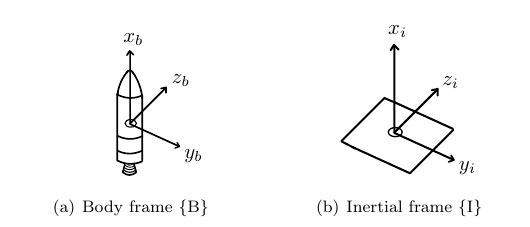}
    \caption{Reference frames.}
    \label{frames}
    \end{figure}

To describe the dynamics and kinematics of the launcher, it is crucial to define the reference frames to be used. Two reference frames are used: a body-fixed one \{B\} (Fig.\ref{frames}a), where the equations of motion are written; and an inertial, space-fixed one \{I\} (Fig.\ref{frames}b).

The coordinate transformation between both reference frames is defined using the Euler angles representation, $\boldsymbol\lambda =\left[\, \phi \:\: \theta \:\: \psi\,\right]^T$, where $\phi$ is the roll angle, $\theta$ is the pitch angle, and $\psi$ is the yaw angle. With this representation, the transformation from \{B\} to \{I\} is obtained through a sequential rotation $\mathbf{R}(\boldsymbol{\lambda})=\mathbf{R}_z(\psi)\,\cdot\,\mathbf{R}_y(\theta)\,\cdot\,\mathbf{R}_x(\phi)$, where $\mathbf{R(\boldsymbol{\lambda})} \in SO\,(3)$ is given by
\begin{equation*}
    \mathbf{R(\boldsymbol{\lambda})} = \begin{bmatrix}
c_{\theta}c_{\psi} & s_{\phi}s_{\theta}c_{\psi} - c_{\phi}s_{\psi} & c_{\phi}s_{\theta}c_{\psi} + s_{\phi}s_{\psi}\\
c_{\theta}s_{\psi} & s_{\phi}s_{\theta}s_{\psi} + c_{\phi}c_{\psi} & c_{\phi}s_{\theta}s_{\psi} - s_{\phi}c_{\psi}\\
-s_{\theta}        & s_{\phi}c_{\theta}                              & c_{\phi}c_{\theta}
\end{bmatrix}\,,
\end{equation*}
in which $c$ and $s$ stand as abbreviations for the trigonometric functions. The inverse transform, from \{I\} to \{B\}, is defined by the transpose $\mathbf{R^T(\boldsymbol{\lambda})}$.

\subsection{Dynamics and kinematics}
Using Newton-Euler's equations for rigid body translational and rotational motion, the dynamics and kinematics of the launcher in the six degrees of freedom are obtained 
\begin{equation}\label{generic_model}
\begin{cases}
  \mathbf{\dot{p}} = \mathbf{R(\boldsymbol{\lambda})}\,\mathbf{v}\\
  \mathbf{\dot{R}(\boldsymbol{\lambda})} = \mathbf{R(\boldsymbol{\lambda})}\,\mathbf{S}(\boldsymbol\omega)\\
  m\,\dot{\mathbf{v}} = -\mathbf{S}(\boldsymbol\omega)\,m\,\mathbf{v} + \mathbf{f}\\
  \mathbf{J}\,\dot{\boldsymbol\omega} = -\mathbf{S}(\boldsymbol\omega)\,\mathbf{J}\,\boldsymbol\omega + \boldsymbol\tau\\
\end{cases},
\end{equation}
where $\mathbf{p}=\left[\,x_i\:\:y_i\:\:z_i\,\right]^T$ is the position in the inertial frame, $\mathbf{v}=\left[\,u\:\:v\:\:w\,\right]^T$ is the velocity expressed in the body frame, $\boldsymbol\omega = \left[\,p\:\:q\:\:r\,\right]^T$ is the angular velocity expressed in the body frame, $m$ is the mass, $\mathbf{S}(.)$ is a skew-symmetric matrix, $\mathbf{f} \in \mathbb{R}^3$ is the external force expressed in the body frame, $\mathbf{J}$ is the inertia matrix, and $\boldsymbol\tau \in \mathbb{R}^3$ is the external torque expressed in the body frame. Following the axial symmetry assumption, the cross-products of inertia can be assumed as zero and the $y$ and $z$ terms can be assumed equal, resulting in a diagonal matrix, $\mathbf{J} = diag\,(J_l,J_t,J_t)\,$,
where $J_l$ denotes the longitudinal inertia and $J_t$ denotes the transverse inertia.

\subsubsection{External forces and torques}

The total external force can be decomposed as $\mathbf{f} = \mathbf{f_g}+\mathbf{f_p}+\mathbf{f_a}$, where $\mathbf{f_g}$ represents the gravity force, $\mathbf{f_p}$ the propulsive force, and $\mathbf{f_a}$ the aerodynamic force, all expressed in \{B\}. As for the external torque, it is given by $\boldsymbol\tau = \boldsymbol{\tau}_\mathbf{p} + \boldsymbol{\tau}_\mathbf{a} + \boldsymbol{\tau}_\mathbf{r} $, where $\boldsymbol{\tau}_\mathbf{p}$ represents the propulsive control torque, $\boldsymbol{\tau}_\mathbf{a}$ represents the aerodynamic torque, and $ \boldsymbol{\tau}_\mathbf{r}$ is the reaction control torque provided by the additional system, all expressed in \{B\}.  
\subsubsection*{Gravitational}
Under the stated assumption, and considering the definition of the inertial frame \{I\}, the gravity force is simply
\begin{equation}\label{gravity}
\mathbf{f_g} = \mathbf{R^T(\boldsymbol{\lambda})}\,\begin{pmatrix}
-mg\\
0\\
0\\
\end{pmatrix} =
\begin{pmatrix}
-mg\, c_{\theta}c_{\psi}\\
-mg\,(s_{\phi}s_{\theta}c_{\psi} - c_{\phi}s_{\psi})\\
-mg\,(c_{\phi}s_{\theta}c_{\psi} + s_{\phi}s_{\psi})
\end{pmatrix},
\end{equation}
where $g$, the gravitational acceleration, varies with the altitude according to $g=g_0\,R_E^2\,/\,(R_E+h)^2$, in which $g_0$ is the gravitational acceleration constant at surface level, $R_E$ is the mean Earth radius, and $h$ is the altitude.
\subsubsection*{Propulsive}
Considering ideal propulsion, and all its underlying assumptions, the thrust force produced by the motor is \cite{prop_sutton}
\begin{equation*}
    T = \underbrace{|\Dot{m}|\cdot v_e}_{\text{Dynamic}} + \underbrace{(p_e - p_a)\cdot A_e}_{\text{Static}}\,,
    \label{thrust}
\end{equation*}
where $\dot{m}$ is the mass flow rate, $v_e$ is the effective exhaust velocity, $p_e$ is the nozzle exit pressure, $p_a$ is the atmospheric pressure, and $A_e$ is the nozzle exit area. Two separate contributions can be identified: the dynamic one, caused by the exhaust of the expanded combustion gases; and the static, caused by the pressure gradient between the nozzle exit and the atmosphere. 

\begin{figure}[!htpb]
    \centering
    \includegraphics{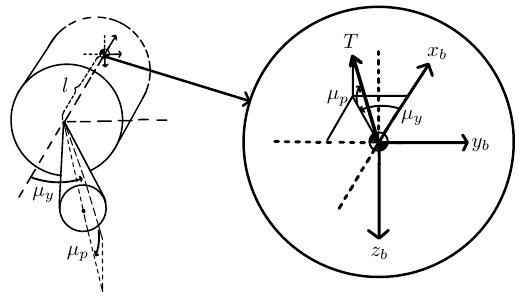}
    \caption{Thrust vector decomposition in the body axes.}
    \label{tvcfig}
    \end{figure}

To obtain the resultant propulsive force and torque, the thrust vector has to be decomposed in the three body axes as illustrated in Fig. \ref{tvcfig}.
According to it, the thrust vector is decomposed using the angles $\mu_p$ and $\mu_y$, which are the control inputs, where $\mu_p$ is the gimbal angle that, on its own, produces a pitching moment, and $\mu_y$ is the one that produces a yawing moment. Using these angles, the propulsive force and torque in the body frame are, respectively,
\begin{equation}\label{tvceqs}
\mathbf{f_p} = 
  \begin{pmatrix}
T\, c_{\mu_p}c_{\mu_y}\\
-T\, c_{\mu_p}s_{\mu_y}\\
-T\, s_{\mu_p}
\end{pmatrix}\,\text{and}\:\: 
\boldsymbol{\tau_p} = \begin{pmatrix}
0\\
-T\, s_{\mu_p}\, l\\
T\, c_{\mu_p}s_{\mu_y}\, l
\end{pmatrix}\,, 
\end{equation}
where $l$ is the control torque arm, which corresponds to the distance between the nozzle gimbal point and the centre of mass of the rocket, $x_{cm}$, measured from the tip of the rocket. 
\subsubsection*{Aerodynamic}

The aerodynamic force and torque, expressed in the body frame, can be modelled as
\begin{equation}\label{aerodynamic}
\mathbf{f_a}  = 
  \begin{pmatrix}
-\overline{q}\, C_A\, S\\
\overline{q}\, C_{Y}\, S\\
-\overline{q}\, C_{N}\, S
\end{pmatrix}\,,\hspace{10pt}\boldsymbol{\tau_a} = 
  \begin{pmatrix}
\overline{q}\, C_l\, S\, d\\
\overline{q}\, C_m\, S\, d\\
\overline{q}\, C_n\, S\, d
\end{pmatrix}\,,       
\end{equation}
where $\overline{q}$ is the dynamic pressure, $d$ is the diameter of the fuselage, $S$ its cross-sectional area, $C_A$, $C_Y$, and $C_N$ are, respectively, the axial, lateral, and normal aerodynamic force coefficients, and $C_l$, $C_m$, and $C_n$ are, respectively, the rolling, pitching, and yawing aerodynamic moment coefficients.

The normal and lateral force coefficients can be determined using a linear relation with the aerodynamic angles of attack, $\alpha$, and side-slip, $\beta$: ${C_Y} = {C_Y}_\beta\, \beta$ and ${C_N} = {C_N}_\alpha\, \alpha$, whose derivatives (${C_Y}_\beta$ and ${C_N}_\alpha$) depend mainly on the angles and Mach number. As for the axial force coefficient, $C_A$, in most applications, its dependency on the aerodynamic angles can be neglected and it is assumed to vary only with Mach number. The relevant velocity for aerodynamic computations is the one expressed in relation to the fluid composing the atmosphere, $\mathbf{v_{\text{rel}}}=\left[\,u_{\text{rel}}\:\:v_{\text{rel}}\:\:w_{\text{rel}}\,\right]^T$. This is given by $\mathbf{v_{\text{rel}}} = \mathbf{v} - \mathbf{v_w}$, where $\mathbf{v_w}$ is the wind velocity vector expressed in \{B\}. The aerodynamic angles are then given by $\alpha = tan^{-1}(w_{\text{rel}}/u_{\text{rel}})$ and $\beta = sin^{-1}(v_{\text{rel}}/V_{\text{rel}})$, where $V_{\text{rel}}$ is the norm of the relative velocity vector.

Regarding the moment coefficients, if the reference moment station is defined as the centre of pressure, and its location, $x_{cp}$, measured from the tip of the rocket, can be determined, the reference moments are zero and the moment coefficients take the form $C_l={C_l}_p\, p\,d/(2V_{\text{rel}})$, $C_m=-{C_N}\,S.M+ (C_{m_{q}}+C_{m_{\dot{\alpha}}})\, q\,d/(2V_{\text{rel}})$, and $C_n=-{C_Y}\,S.M + (C_{n_{r}}+C_{n_{\dot{\beta}}})\, r\,d/(2V_{\text{rel}})$, where the static stability margin, $S.M = (x_{cp} - x_{cm})/d$, intuitively appears, and $C_{l_p}$, $C_{m_q}$, $C_{m_{\dot{\alpha}}}$, $C_{n_r}$, and $C_{n_{\dot{\beta}}}$ are all aerodynamic damping coefficients. 
\subsubsection{Explicit dynamics and kinematics}
The explicit dynamics and kinematics can be retrieved by substituting the total external force and torque in (\ref{generic_model}) by all the individual detailed components, (\ref{gravity}), (\ref{tvceqs}), and (\ref{aerodynamic}), yielding
\begin{equation}\label{model}
\begin{cases}
  \dot{u} = -g\,c_{\theta}c_{\psi} -\frac{\overline{q}}{m}\,S\,C_A + \frac{T}{m}c_{\mu_1}\,c_{\mu_2} -q\,w + r\,v\\[0.1cm]
  \dot{v} = -g\,(s_{\phi}s_{\theta}c_{\psi}-c_{\phi}s_{\psi})\! + \frac{\overline{q}}{m}SC_{Y} - \frac{T}{m}c_{\mu_1}s_{\mu_2} - r\,u+\!p\,w \\[0.1cm]
  \dot{w} = -g\,(c_{\phi}s_{\theta}c_{\psi}+s_{\phi}s_{\psi}) - \frac{\overline{q}}{m}\,S\,C_{N} - \frac{T}{m}\,s_{\mu_1}-p\,v + q\,u\\[0.1cm]
  \dot{p} = {J_l}^{-1}\,(\,\overline{q}\,S\,d\,C_l + \tau_r\,)\\[0.1cm]
  \dot{q} = {J_t}^{-1}\,(\,\overline{q}\,S\,d\,C_m - T\,s_{\mu_1}\,l\,)\\[0.1cm]
    \dot{r} = {J_t}^{-1}\,(\,\overline{q}\,S\,d\,C_n + T\,c_{\mu_1}\,s_{\mu_2}\,l\,)\\[0.1cm]
    \dot{\phi} =p+(q\,s_{\phi}+r\,c_{\phi})\,t_{\theta}\\[0.1cm]
     \dot{\theta} =q\,c_{\phi} - r\,s_{\phi}\\[0.1cm]
     \dot{\psi} = \displaystyle\frac{q\,s_{\phi} + r\,c_{\phi}}{c_{\theta}}\\
\end{cases}
\end{equation}
It is noted that by using the Euler angles representation a singularity arises for $\theta = \pm\frac{\pi}{2}$, however, the way the reference frames are defined prevents the rocket to reach this attitude inside the admissible range of operation (far from horizontal orientation).

\section{Linearized physical model}
\label{sec:linearss}

Linear control and estimation techniques, such as the LQR and the Kalman filter, rely on mathematical representations of the linear systems under study. These representations are usually written in the state-space form. In this Section, a generic state-space model for a thrust vector controlled launch vehicle is obtained by linearizing the already detailed explicit dynamics and kinematics in (\ref{model}). 

A widely used linearization technique consists in finding an equilibrium point of the system, in which the first-order derivatives of the states are null, and performing a Taylor series expansion, considering small perturbations around the equilibrium condition. However, rocket flight is dominated by highly varying conditions and parameters, such as mass and inertia, aerodynamic coefficients, dynamic pressure, and thrust, which make it impossible to find a so called trimming trajectory, for which equilibrium is reached with constant control inputs. 

One viable alternative \citep{Tewari}, is to linearize the system at multiple points, denominated as operating points, throughout a previously selected reference trajectory. The selected trajectory will impose the reference values for system states ($x_0$) and inputs ($u_0$), and the outcome is a linear time-varying system. Linear controllers can be designed for the state-space representations associated with each operating point and then scheduled during flight. Therefore, the operating points have to be selected so as to capture all the relevant dynamics of the system, preventing that the system destabilizes.

The Taylor series expansion is still used, but now at each operating point. Firstly, the following variable transformations are defined: $\delta x = x-x_0$ and $\delta u = u-u_0$;  where $\delta\,x$ and $\delta u$ are small perturbations around the reference values for each point. By using the variable transformation in the non-linear differential equations of the system (denoted by $\dot{x} = f(x,u)$), generically, we have that $\delta\dot{x} =  f(x,u) - f(x_0,u_0) = f(x_0 + \delta x, u_0 + \delta u) - f(x_0,u_0)$. Using the Taylor series expansion of $f(x_0 + \delta x, u_0 + \delta u)$ around ($x_0$, $u_0$), and neglecting the higher-order terms, we obtain 
   \begin{equation*}
   \delta \dot{x} = f(x_0,u_0) + \frac{\partial f}{\partial x}\bigg|_{x_0, u_0}\cdot\delta x\, + \frac{\partial f}{\partial u}\bigg|_{x_0, u_0}\cdot\delta u\,  - f(x_0,u_0)\,,
 \end{equation*} 
which simplifies to 
   \begin{equation}\label{linearizer}
   \delta \dot{x}=\frac{\partial f}{\partial x}\bigg|_{x_0, u_0}\cdot\delta x\, + \frac{\partial f}{\partial u}\bigg|_{x_0, u_0}\cdot\delta u\,.
   \end{equation}

Expression (\ref{linearizer}) is then applied to all non-linear first order differential equations in (\ref{model}), yet with further simplifications: the roll rate ($p$) is assumed to be null; the roll angle ($\phi$) is taken as constant parameter rather than a state; wind velocity is considered to be zero; the actuator dynamics are not included in the model; and system parameters are considered constant at each operating point (frozen parameters). The first two simplifications are due to the fact that roll control is achieved by an additional system, the third one makes the relative velocity vector equal to the linear velocity vector expressed in the body frame, and the final one removes the existent dependencies of the parameters on the state variables when computing the Taylor derivatives.

Considering a generic reference trajectory, the resultant state-space representation follows
\begin{subequations}
\label{ss}
\begin{equation}
    \delta \mathbf{x} = \left[\,\delta u\:\: \delta v\:\: \delta w\:\: \delta q\:\: \delta r\:\: \delta \theta\:\: \delta \psi\,\right]^{T}\,,\hspace{10pt}
    \delta \mathbf{u} = \left[\,\delta \mu_p\:\: \delta \mu_y\,\right]^{T}\,,
\end{equation}
\begin{equation}
 \delta \mathbf{\dot{x}}(t) = \mathbf{A}(t)\cdot\delta \mathbf{x}(t) + \mathbf{B}(t)\cdot\delta \mathbf{u}(t)\,,
\end{equation}
\begin{equation}
     \renewcommand*{\arraystretch}{1.2}
	 \mathbf{A}(t) = \begin{bmatrix}
			0 & r_0 & -q_0 & -w_0 & v_0 & a_{16} & a_{17}\\
			-r_0 & a_{22} & 0 & 0 & -u_0 & a_{26} & a_{27}\\
	       a_{31} & 0 & a_{33} & u_0 & 0 & a_{36} & a_{37}\\
           a_{41} & 0 & a_{43} & a_{44} & 0 & 0 & 0\\
           	0 & a_{52} & 0 & 0 & a_{55} & 0 & 0\\
			0 & 0 & 0 & c_{\phi_0} & -s_{\phi_0} & 0 & 0\\
			0 & 0 & 0 & s_{\phi_0}/c_{\theta_0} & a_{85} & a_{86} & 0
		\end{bmatrix}\,,
\end{equation}
\begin{equation}
\renewcommand*{\arraystretch}{1.2}
	\mathbf{B}(t) = \begin{bmatrix}
	        -\frac{T}{m}\,s_{\mu_{1_0}}\,c_{\mu_{2_0}} & -\frac{T}{m}\,c_{\mu_{1_0}}\,s_{\mu_{2_0}} \\[0.2cm]
	        \frac{T}{m}\,s_{\mu_{1_0}}\,s_{\mu_{2_0}} & -\frac{T}{m}\,c_{\mu_{1_0}}\,c_{\mu_{2_0}}\\[0.2cm]
	        -\frac{T}{m}\,c_{\mu_{1_0}} & 0\\[0.2cm]
	        -\frac{T\,l}{J_t}\,c_{\mu_{1_0}} & 0\\[0.2cm]
	        -\frac{T\,l}{J_t}\,s_{\mu_{1_0}}\,s_{\mu_{2_0}} & \frac{T\,l}{J_t}\,c_{\mu_{1_0}}\,c_{\mu_{2_0}}\\[0.2cm]
	        0 & 0\\
	        0 & 0
	\end{bmatrix}\,,
	\end{equation}
\end{subequations}
with
\begin{equation*}
  \begin{split}
    a_{16} &= g\,s_{\theta_0}\,c_{\psi_0}\\
    a_{17} &= g\,c_{\theta_0}\,s_{\psi_0}\\
    a_{22} &= \frac{\overline{q}\,S\,C_{Y_{\beta}}}{m\,{\left(1-\frac{v_0^2}{{V_0}^2}\right)}^{1/2}\,V_0}\\
    a_{26} &= -g\,(s_{\phi_0}\,c_{\theta_0}\,c_{\psi_0}-c_{\phi_0}\,s_{\psi_0})\\
    a_{27} &= g\,(s_{\phi_0}\,s_{\theta_0}\,s_{\psi_0}+c_{\phi_0}\,c_{\psi_0})\\[0.1cm]
    a_{31} &= q_0 + \frac{\overline{q}\,S\,C_{N_{\alpha}}\,w_0}{m\,({u_0}^2+{w_0}^2)}\\[0.2cm]
    a_{33} &= -\frac{\overline{q}\,S\,C_{N_{\alpha}}\,u_0}{m\,({u_0}^2+{w_0}^2)}\\
    a_{36} &= -g\,c_{\phi_0}\,c_{\theta_0}\,c_{\psi_0}\\
    a_{37} &= -g\,(-c_{\phi_0}\,s_{\theta_0}\,s_{\psi}+s_{\phi_0}\,c_{\psi_0})\\
  \end{split}
\quad
  \begin{split}
    a_{41} &= \frac{\overline{q}\,S\,d\,SM\,C_{N_{\alpha}}\,w_0}{J_t\,({u_0}^2+{w_0}^2)}\\[0.2cm]
    a_{43} &= -\frac{\overline{q}\,S\,d\,SM\,C_{N_{\alpha}}\,u_0}{J_t\,({u_0}^2+{w_0}^2)}\\[0.2cm]
    a_{44} &= \frac{\overline{q}\,S\,{d}^2\,(C_{m_{q}} + C_{m_{\dot{\alpha}}})}{2\,J_t\,V_0}\\[0.2cm]
    a_{52} &= -\frac{\overline{q}\,S\,d\,SM\,C_{Y_{\beta}}}{J_t\,V_0\,{\left(1-\frac{v_0^2}{{V_0}^2}\right)}^{1/2}}\\[0.2cm]
    a_{55} &= \frac{\overline{q}\,S\,{d}^2\,(C_{n_{r}} + C_{n_{\dot{\beta}}})}{2\,J_t\,V_0}\\
    a_{85} &= \frac{c_{\phi_0}}{c_{\theta_0}}\\[0.1cm]
    a_{86} &= \frac{(q_0\,s_{\phi_0}+r_0\,c_{\phi_0})\,s_{\theta_0}}{{{c^2}_{\theta_0}}}\\
  \end{split}
\end{equation*}
where $\mathbf{A}(t)$ and $\mathbf{B}(t)$ are the state-space matrices given by the first-order Taylor derivatives in (\ref{linearizer}) with respect to system states and inputs, respectively, calculated at the operating points, and $V_0$ is the norm of $\mathbf{v}$ at each operating point. Due to the aforementioned simplifications, $p$ and $\phi$ are not states of the system, even though they are physical variables in the complete non-linear model.

\section{Architecture}
\label{sec:arch}
\begin{figure*}
    \centering
    \includegraphics{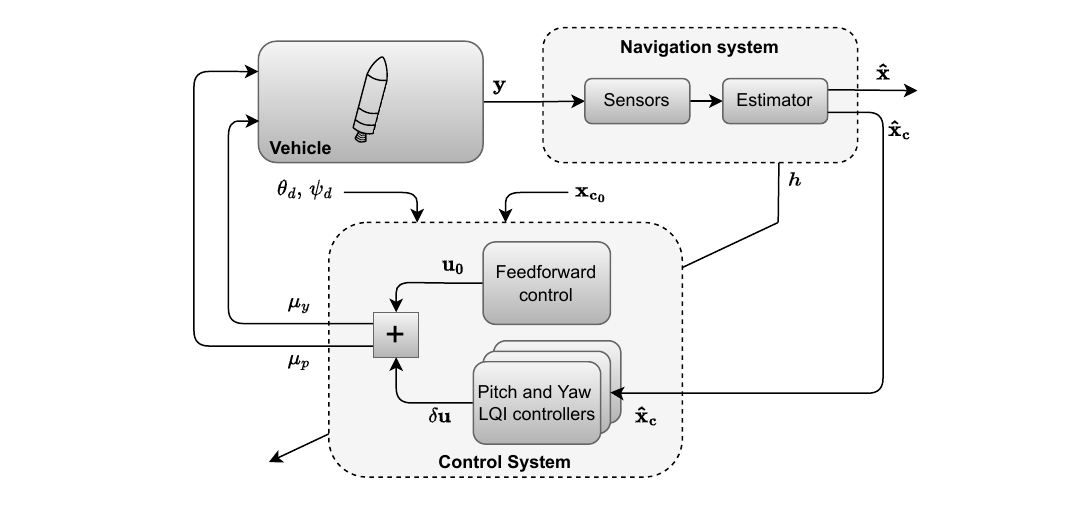}
    \caption{System architecture.}
    \label{arch}
    \end{figure*}

To achieve a stable solution with accurate reference tracking for the pitch and yaw angles of a naturally unstable launcher, the integrated architecture in Fig. \ref{arch}, comprising both the navigation and control systems, is proposed. The underlying principles are LQR control with integrative components (LQI) and linear Kalman filtering.
    
The navigation system is composed by sensors, which measure relevant quantities associated with rocket flight, $\mathbf{y}$), and an estimator, based on Kalman filtering, which provides estimates on the state vector, $\mathbf{\hat{x}}$, given the sensors' readings, with the added benefit of filtering the sensors' noise and correcting its bias. The subset of the estimated state vector, $\mathbf{\hat{x}}$, used for feedback control is represented by $\mathbf{\hat{x}_c}$.

The control system is divided in two major blocks: feedforward control and LQI feedback control. Feedforward control consists in the pre-determined values for the system inputs, $u_0$, that allow the vehicle to follow the reference trajectory under nominal conditions, i.e, without perturbations and model uncertainties. On the other hand, feedback LQI control is responsible for ensuring stability and accurate reference tracking ($\theta_d$ and $\psi_d$) in a real flight scenario. 

Feedback control is implemented in the perturbation domain, meaning that the reference values of the states used for feedback, $\mathbf{x_{c_0}}$, are needed to retrieve the perturbed states according to $\delta \mathbf{\hat{x}_c} = \mathbf{\hat{x}_c} - \mathbf{x_{c_0}}$. It acts on the perturbed states using the optimal gains calculated for each operating point through the use of the LQI control law and the respective state-space representation. To ensure a smooth time evolution in the control inputs, linear interpolation is used to schedule the gains. The variable selected to interpolate the gains is the altitude, $h$, to avoid potential mismatches resulting from delays that could occur in a time-based interpolation. The scheduled controllers are represented in the Fig. \ref{arch} by multiple block layers.

By summing the feedforward and feedback control values, respectively $\mathbf{u_0}$ and $\delta \mathbf{u}$, the control inputs, $\mu_p$ and $\mu_y$, are obtained.

\section{Navigation}
\label{sec:navigation}
As mentioned, the navigation system is composed by sensors and an estimator. In this section, the selected sensor suite, as well as the estimator design, are detailed. 
\subsection{Sensor suite}
To design a navigation system, it is necessary to select the sensor suite that will be on board of the vehicle. Sensors might provide a direct measurement on the required state variables or on other quantities that can then be used to estimate them. For launch vehicles, and taking into account the state variables to be measured - position, linear and angular velocities, and Euler angles - it is common to use an Inertial Measurement Unit (IMU) combined with a Global Navigation Satellite System (GNSS) receiver. If not included in the IMU, barometers and magnetometers are also standard.

The IMU is composed by 3-axis accelerometers and gyroscopes. An accelerometer supplies a measure of the system's acceleration and can be used to determine the vehicle's velocity by integration. To do so, it is necessary to know the initial condition. Over time, the velocity measurement will drift from the true value due to the inherent noise and bias properties of the accelerometer. By combining the 3-axis accelerometers, a measurement on the linear acceleration vector in the body frame is obtained, $\mathbf{a_r} \in \mathbb{R}^3$.

A gyroscope provides a measurement of the system's angular rate. The angular rate measurements, $\boldsymbol{\omega_r}\in \mathbb{R}^3$, can be integrated to determine an estimate of the system's attitude. Once again, the calculated attitude drifts boundlessly from the true attitude of the system due to the inherent noise and bias properties of the gyroscope.

If the 3-axis accelerometer is assumed to be measuring gravity alone, it is possible to calculate the pitch and yaw angles from the direction of the gravity vector. However, since the accelerometer is assumed to be measuring gravity alone, any added dynamic motion causes an error in the calculation of the system's pitch and yaw. A magnetometer can be used to obtain a measurement of the roll angle by comparing the measurement of the magnetic field surrounding the system to Earth's magnetic field. The combined attitude solution is $\boldsymbol{\lambda_r} = \left[\,\phi_r\:\:\theta_r\:\:\psi_r\,\right]^T$.

A GNSS is a satellite configuration, or constellation, that provides coded satellite signals which are processed by a GNSS receiver inside the vehicle to calculate position, velocity, and time. In this paper, the position measurements by the GNSS receiver, $\mathbf{p_r} \in \mathbb{R}^3$, are assumed to be already translated into the inertial frame. Additionally, the velocity given by the GNSS receiver can be used to remove the dynamic acceleration component in the accelerometer readings when computing the pitch and yaw angles. 

\subsection{Estimator design}

In this subsection, the estimator design is presented, with its overall architecture and the individual components, based on Kalman filtering, being detailed.

\subsubsection{Estimator architecture}

\begin{figure}[h]
    \centering
    \includegraphics{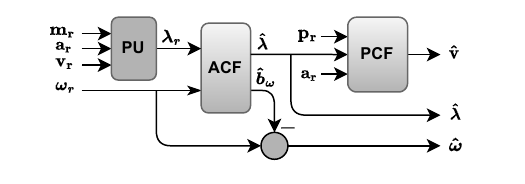}
    \caption{Estimator architecture.}
    \label{estimatorarch}
    \end{figure}

The estimator is composed by two complementary filters and a pre-processing unit, according to the scheme in Fig. \ref{estimatorarch}.
The pre-processing unit (PU) takes the magnetometer, $\mathbf{m_r}$, accelerometer, $\mathbf{a_r}$, and GNSS velocity, $\mathbf{v_r}$, readings to obtain an indirect measurement on the Euler angles, $\boldsymbol{\lambda_r}$. The underlying algorithms are widely available \citep{celis,batista} and are not here presented. 

The first filter is the Attitude Complementary Filter (ACF), which uses the Euler angles readings, $\boldsymbol\lambda_r$, and the measured angular rates from the gyroscopes, $\boldsymbol\omega_r$, to provide a filtered attitude estimate, $\hat{\boldsymbol\lambda}$, and an estimate on the bias of the three angular rates, $\mathbf{b_{\boldsymbol{\omega}}} \in \mathbb{R}^3$, to correct the signal from the sensor.

The second one is the Position Complementary Filter (PCF), which merges the position readings from the GNSS receiver, translated into the inertial frame, $\mathbf{p_r}$, and the acceleration measurements from the accelerometer, $\mathbf{a_r}$, to provide an estimate on the linear velocity vector, $\hat{\mathbf{v}}$. This filter is also self-calibrated since it accounts for the bias in the three acceleration readings, $\mathbf{b_a} \in \mathbb{R}^3$. 

\subsubsection{Kalman filter}
The Kalman filter is a widely used observer to tackle the estimation problem for linear dynamic systems \cite{simon}. When both the process and measurement associated with the estimated state are corrupted by random, independent, zero mean Gaussian white noise, the solution provided by the Kalman filter is statistically optimal with respect to any quadratic function of the estimation error. For this reason, it is also referred to as Linear Quadratic Estimator (LQE), and represents the dual of the LQR to the estimation problem. 

In continuous time, the random process and observation are given by
\begin{equation*}
\begin{cases}
 \dot{\mathbf{x}} = \mathbf{A}\,\mathbf{x}+\mathbf{B}\,\mathbf{u}+\mathbf{G}\,\mathbf{w}\\[0.1cm]
 \mathbf{y} = \mathbf{C}\,\mathbf{x} + \mathbf{v}\\
 \end{cases},
\end{equation*}
where all terms are time dependant, $\mathbf{w}$ is the process noise (associated with the model), $\mathbf{v}$ is the measurement noise (associated with the sensors) and $\mathbf{G}$ is the process noise coupling matrix. These random noises are represented by the covariance matrices $\mathbf{Q}$ and $\mathbf{R}$ for the process and measurement noise, respectively. The $\mathbf{Q}$ and $\mathbf{R}$ matrices are positive semi-definite.

Given the defined process, observation and noise properties, the Kalman filter is capable of providing an optimal state estimation according to the differential equation
\begin{equation*}
 \dot{\hat{\mathbf{x}}} = \mathbf{A}\,\hat{\mathbf{x}} + \mathbf{B}\,\mathbf{u} + \mathbf{L}\,\left(\mathbf{y} - \mathbf{C}\,\hat{\mathbf{x}}\right)\,,
\end{equation*} 
in which $\hat{\mathbf{x}}$ is the state estimate and $\mathbf{L}$ is the Kalman gain. Given an initial condition $\hat{\mathbf{x}}(0)$, the state estimate derivative $\dot{\hat{\mathbf{x}}}$ is recursively propagated by correcting the process with the state estimation error ($\mathbf{y}-\mathbf{C}\,\hat{\mathbf{x}}$) multiplied by the Kalman gain. The Kalman gain is given by
\begin{equation*}
 \mathbf{L} = \mathbf{P}\,\mathbf{C}^T\,\mathbf{R}^{-1}\,,
\end{equation*}
where $\mathbf{P}$ is the solution to the matrix Riccati differential equation
\begin{equation}\label{matrixriccati}
\dot{\mathbf{P}} = \mathbf{A}\,\mathbf{P} + \mathbf{P}\,\mathbf{A}^T + \mathbf{G}\,\mathbf{Q}\,\mathbf{G}^T - \mathbf{P}\,\mathbf{C}^T\,\mathbf{R}^{-1}\,\mathbf{C}\,\mathbf{P}\,.
\end{equation}
If the process is time-varying, this equation has to be continuously solved. However, for the steady-state case, $\dot{\mathbf{P}}$ is zero and (\ref{matrixriccati}) simplifies to the famous Algebraic Riccati Equation (ARE). For the ARE to have a unique positive definite solution $\mathbf{P}$, it is a sufficient condition that the pair ($\mathbf{A}$, $\mathbf{C}$) is observable.

The tuning parameters will be the $\mathbf{Q}$ and $\mathbf{R}$ noise covariance matrices. The $\mathbf{R}$ matrix can be tuned according to the specifications of the on-board sensors, while determining the model noise covariance, $\mathbf{Q}$, might represent a harder task. Resorting to simulation in order to properly tune the $\mathbf{Q}$ matrix is a good initial method, which can later be updated using results coming from a real implementation scenario.
\subsubsection{ACF}
For the ACF, it is assumed that the Euler angles measurement is corrupted by Gaussian white-noise, $\boldsymbol{w_\lambda}$, as well as the angular rates readings, $\boldsymbol{w_\omega}$, and that the gyroscope bias is driven by a Gaussian noise sequence, $\boldsymbol{n_{b_\omega}}$, yielding
\begin{equation*}
\boldsymbol{\lambda_r} = \boldsymbol\lambda + \boldsymbol{w_\lambda}\,,
\end{equation*}
\begin{equation*}
\boldsymbol{\omega_r} = \boldsymbol\omega + \boldsymbol{w_\omega} + \boldsymbol{b_\omega}\,,
\end{equation*}
\begin{equation*}
\boldsymbol{\dot{b}_\omega} := \boldsymbol{n_{b_\omega}}\,,\, \text{with}\,\: \boldsymbol{b_{\omega_0}} = \boldsymbol{\overline{b}_\omega}\,,   
\end{equation*}
where $\boldsymbol{\overline{b}_\omega}$ is an unknown constant offset.

This filter follows the methodology of the Kalman filter, with the process being based on the kinematic equations for the Euler angles presented in (\ref{model}). Furthermore, it uses the Euler
angles readings, $\boldsymbol{\lambda_r}$, in the process matrix, so that the system can be regarded as linear and the derived Kalman theory can be applied. The state-space representation of the observation process follows,
\begin{equation*}
\mathbf{x}_{\text{acf}}=\left[\begin{matrix}
\boldsymbol\lambda & \boldsymbol{b_\omega}
\end{matrix}\right]^T\,,\hspace{10pt}\mathbf{y}_{\text{acf}} = \boldsymbol{\lambda_r} + \boldsymbol{w_\lambda}\,,\hspace{10pt} \mathbf{\hat{y}}_{\text{acf}} = \boldsymbol{\hat{\lambda}}\,,
\end{equation*}
\begin{equation*}
 \renewcommand*{\arraystretch}{1.2}
\mathbf{\dot{\hat{x}}}_{\text{acf}} =
\begin{bmatrix}
\mathbf{0_{3x3}} & \mathbf{A}_{\text{acf}}\\
\mathbf{0_{3x3}} & \mathbf{0_{3x3}}
\end{bmatrix}\mathbf{\hat{x}}_{\text{acf}}+
\begin{bmatrix}
\mathbf{B}_{\text{acf}}\\
\mathbf{0_{3x3}}
\end{bmatrix}\boldsymbol{\omega_r} +
\mathbf{L}_{\text{acf}}\,(\mathbf{y}_{\text{acf}}-\mathbf{\hat{y}}_{\text{acf}})\,,
\end{equation*}
with
\begin{equation*}
\renewcommand*{\arraystretch}{1.2}
\mathbf{A}_{\text{acf}} =
\begin{bmatrix}
 -1 & -s_{\phi_r}\,t_{\theta_r} & -c_{\phi_r}\,\,t_{\theta_r}\\
0 & -c_{\phi_r} & s_{\phi_r}\\
0 & -\displaystyle\frac{s_{\phi_r}}{c_{\theta_r}} & -\displaystyle\frac{c_{\phi_r}}{c_{\theta_r}}
\end{bmatrix}\,,\hspace{10pt}
\mathbf{B}_{\text{acf}} = -\mathbf{A}_{\text{acf}}\,.
\end{equation*}
To calculate the gain matrix $\mathbf{L}_{\text{acf}}$, of dimension 6 by 3, the time-invariant equivalent of the system is obtained by choosing the vertical attitude, $\boldsymbol\lambda_r=\left[0\:\:0\:\:0\right]^T$, to define the process matrices and compute the time-invariant Kalman gains.
\subsubsection{PCF}
For the PCF, both the position and acceleration measurements are considered to be corrupted by Gaussian white noise, $\mathbf{w_p}$ and $\mathbf{w_a}$, and the accelerometer bias is also driven by a Gaussian noise sequence, $\mathbf{n_{b_a}}$, yielding
\begin{equation*}
\mathbf{a_r} = \mathbf{a} + \mathbf{w_a} + \mathbf{b_a}\,,
\end{equation*}
\begin{equation*}
\mathbf{\dot{b}_a} := \mathbf{n_{b_a}}\,,\, \text{with}\,\: \mathbf{b_{a_0}} = \mathbf{\overline{b}_a}\,,   
\end{equation*}
where $\mathbf{\overline{b}_a}$ is an unknown constant offset.

This filter is also kinematic and follows the Kalman filter formulation, considering the following equations of motion,
\begin{equation*} 
\dot{\mathbf{p}} = \mathbf{R(\boldsymbol{\lambda})}\,{\mathbf{v}}\,,\hspace{10pt}
\mathbf{\ddot{p}} = \mathbf{R(\boldsymbol{\lambda})}\,{\mathbf{a}}\,, 
\end{equation*}
where $\mathbf{p}$, as before, is the position in the inertial frame and $\mathbf{a}$ is the acceleration expressed in the body frame. The state-space representation of the filter is then obtained,
\begin{equation*}
\mathbf{x}_{\text{pcf}}=\left[\begin{matrix}
\mathbf{p} & \mathbf{\dot{p}} & \mathbf{b_a}
\end{matrix}\right]^T\,,\hspace{10pt}\mathbf{y}_{\text{pcf}} = \mathbf{p_r} + \mathbf{w_p}\,,\hspace{10pt} \mathbf{\hat{y}}_{\text{pcf}} = \mathbf{\hat{p}}\,,
\end{equation*}
\begin{equation*}
\mathbf{\dot{\hat{x}}}_{\text{pcf}} =
\begin{bmatrix}
\mathbf{0_{3x3}} & \mathbf{I_{3}} & \mathbf{0_{3x3}}\\
\mathbf{0_{3x3}} & \mathbf{0_{3x3}} & -\mathbf{R}\\
\mathbf{0_{3x3}} & \mathbf{0_{3x3}} & \mathbf{0_{3x3}}
\end{bmatrix}\mathbf{\hat{x}}_{\text{pcf}}+
\begin{bmatrix}
\mathbf{0_{3x3}}\\
\mathbf{R}\\
\mathbf{0_{3x3}}
\end{bmatrix}\mathbf{a_r} +
\begin{bmatrix}
\mathbf{L_1}\\
\mathbf{L_2}\\
\mathbf{R}^T\,\mathbf{L_3}
\end{bmatrix}(\mathbf{y}_{\text{pcf}}-\mathbf{\hat{y}}_{\text{pcf}}).
\end{equation*}
The rotation matrix is calculated using the Euler angles estimate from the ACF, i.e, $\mathbf{R(\boldsymbol{\hat{\lambda}})}$. The individual gain matrices $\mathbf{L_1}$, $\mathbf{L_2}$ and $\mathbf{L_3}$, each with dimension 3 by 3, can once again be computed considering the vertical attitude time-invariant, $\boldsymbol{\hat{\lambda}} = \left[0\:\:0\:\:0\right]^T$, to define the rotation matrix, yielding time-invariant Kalman gains. Note that the gain matrix $\mathbf{L_3}$ is associated with the bias vector estimate $\mathbf{\hat{b}_a}$, which is expressed in the body frame, and so it has to be rotated from \{I\} to \{B\}.

\section{Control}
\label{sec:control}
In this section, the feedforward and feedback control components are described. For feedback control, an LQR with integrative action (LQI) is proposed and it is then particularized in a decoupled version.

\subsection{Feedforward control}

Given the natural instability of the system and its time-varying nature, finding the time evolution of the nominal control inputs, $\mathbf{u_0}$, that places the vehicle in the desired trajectory can be a difficult task.

 A first approach could be to solve the non-linear differential equations of the system (\ref{model}) over time such that the attitude reference is correctly followed.  However, this is a mathematically complex problem that would require a numerical solution. 
 
 A more practical strategy is to rely on a simulation model, based on the detailed physical model (\ref{model}), and use a controller that stabilizes the plant and ensures that the reference trajectory is followed in simulation. The resultant actuation values can then be stored to later use in real-time as feedforward control. As long as the model is sufficiently accurate and the varying parameters are approximately known, this approach can be valid.
 
Since the simulated flight is disturbance-free and no uncertainties are added to the model, a simple PID controller per degree of freedom (pitch and yaw), with constant gains, can achieve this task.

\subsection{Feedback control}
\label{feedforward}
Feedback control uses a subset of the state estimates from the navigation system, $\mathbf{\hat{x}_c}$, to stabilize the plant and provide reference tracking of the desired pitch and yaw angles, $\theta_d$ and $\psi_d$. Given the nature of the TVC actuation, trying to control the linear velocities would conflict with the attitude control, specially for non-zero attitude references. Therefore, $\mathbf{x_c} = \left[\,q\:\: r\:\: \theta\:\: \psi\,\right]^{T}$.

\subsubsection{LQR}
The LQR is an optimal controller for linear systems that finds the gain matrix $\mathbf{k}$ in the linear control law $\mathbf{u} = -\mathbf{K}\,\mathbf{x}$, which minimizes a given quadratic cost function formulated as 
\begin{equation*}
    J =  \int_{t}^{T} [\,\mathbf{x}'(\tau)\,\mathbf{Q}\,\mathbf{x}(\tau) + \mathbf{u}'(\tau)\,\mathbf{R}\,\mathbf{u}(\tau)\,] \,d\tau\,,
\end{equation*}
where $\mathbf{Q}$ is a positive semi-definite matrix and $\mathbf{R}$ is a positive definite matrix. In the cost function $J$, the quadratic form $\mathbf{x'\mathbf{Q}x}$ represents a penalty on the deviation of the state $\mathbf{x}$ from the origin, and the term $\mathbf{u'\mathbf{R}u}$ represents the cost of control, making $\mathbf{Q}$ and $\mathbf{R}$ the tuning parameters for the resultant controller. 

It can be shown \cite{friedland} that for the infinite-horizon, or steady-state, version ($T=\infty$), the solution to this optimization problem, which guarantees closed-loop asymptotic stability, is the constant gain matrix
\begin{equation*}
\mathbf{K} = \mathbf{R}^{-1}\,\mathbf{B}^T\,\mathbf{M}\,,
\end{equation*}
 where $\mathbf{M}$ is the solution to the ARE, now formulated for the LQR version,
 \begin{equation}\label{are1}
 \mathbf{M}\,\mathbf{A} + \mathbf{A}^{T}\,\mathbf{M}-\mathbf{M}\,\mathbf{B}\,\mathbf{R}^{-1}\,\mathbf{B}^T\,\mathbf{M} + \mathbf{Q}=\mathbf{0}\,.
 \end{equation}
In order for the ARE (\ref{are1}) to have a unique, positive definite solution $\mathbf{M}$, it is a sufficient condition that the system defined by the pair ($\mathbf{A}$, $\mathbf{B}$) is controllable.

\subsubsection{LQR with integrative component (LQI)}

The LQR feedback control law, applied to the system under study, would ideally drive the states in the perturbation domain to zero, ensuring that the nominal values throughout the trajectory were followed. However, it does not guarantee a zero tracking error for non-zero attitude references ($\theta_d$ and $\psi_d$). In order to have no reference tracking error, and to increase the robustness of the controller to model uncertainties and external perturbations, an integrative component that acts on the attitude tracking error is added.

\begin{figure}[h]
    \centering
    \includegraphics{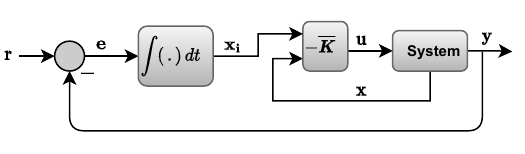}
    \caption{Generic LQI control scheme.}
    \label{genlqi}
    \end{figure}

To obtain this controller using the LQR calculations already detailed, it is only necessary to modify the state-space model when calculating the LQR gains. Generically, the closed-loop control with LQI follows the scheme in Fig. \ref{genlqi}.
Let the difference between the reference signal, $\mathbf{r}$, and the output of the system, $\mathbf{y}$, (the tracking error) be the time derivative of the state-space variables that result from adding the integrative component, $\mathbf{x_i}$. The state-space representation of the resulting regulator can be obtained by combining the open-loop state-space representation with the feedback law,
\begin{equation*}\label{lqieqs}
\mathbf{\dot{z}} = \left(\begin{bmatrix}
                     \mathbf{A} & \mathbf{0}\\
                     -\mathbf{C} & \mathbf{0}
                    \end{bmatrix}-\begin{bmatrix}
                     \mathbf{B}\\
                     \mathbf{0}
                    \end{bmatrix}\,\mathbf{\overline{K}}\right)\mathbf{z} + \begin{bmatrix} \mathbf{0} \\ 1    
\end{bmatrix}\,\mathbf{r}\,,
\end{equation*}
where $\mathbf{z} =\left[\,\mathbf{x}\,\:\mathbf{x_i}\,\right]^T$ is the augmented state vector and $\mathbf{C}$ is the output matrix that selects the output of the system, i.e, the states for reference tracking, from the original state vector ($\mathbf{y} = \mathbf{C}\,\mathbf{x}$). The optimal gain is $\mathbf{\overline{K}} = \left[\,\mathbf{K}\:\:\mathbf{K_i}\,\right]^T$, where $\mathbf{K}$ is the original gain matrix for the state variables, and $\mathbf{K_i}$ is the gain matrix for the integrative components, and can be obtained by solving the ARE using the rearranged state-space matrices
\begin{equation*}
 \mathbf{\overline{A}} = \begin{bmatrix}
                     \mathbf{A} & \mathbf{0}\\
                     -\mathbf{C} & \mathbf{0}
                    \end{bmatrix}\,,\hspace{10pt} \mathbf{\overline{B}} = \begin{bmatrix}       \mathbf{B}\\
                     \mathbf{0}
                    \end{bmatrix}\,.
\end{equation*}

Since the system under study is time-varying, the ARE has to be solved for models coming from each linearization point, resulting in a set of gain matrices to be selected, or scheduled, throughout the flight. Moreover, it is important to note that the state-space representation obtained is expressed in the perturbation domain. The augmented state-vector is
\begin{equation*}
\delta \mathbf{z} = \left[\,\delta u\:\: \delta v\:\: \delta w\:\: \delta q\:\: \delta r\:\: \delta \theta\:\: \delta \psi\:\: \delta \theta_i\:\: \delta \psi_i\,\right]^{T}\,,
\end{equation*}
where $\delta \theta_i$ and $\delta \psi_i$ are the states associated with the integrative components. The $\mathbf{C}$ matrix is given by $\mathbf{C} = \left[\,0\:\:0\:\:0\:\:0\:\:0\:\:1\:\:1\,\right]$, in order to select $\delta \theta$ and $\delta \psi$ as the variables for reference tracking. 

\begin{figure}[h]
    \centering
    \includegraphics{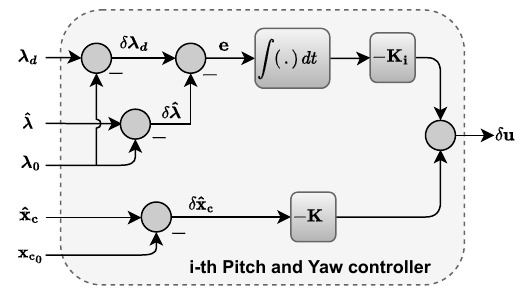}
    \caption{Pitch and yaw LQI controller.}
    \label{controlsys}
    \end{figure}

Given the order of the augmented system and the number of inputs, each gain matrix $\mathbf{\overline{K}}$ will be of dimension 2 by 9, however, since partial feedback is used, $\delta \mathbf{z_c} = \left[\,\delta q\:\: \delta r\:\: \delta \theta\:\: \delta \psi\:\: \delta \theta_i\:\: \delta \psi_i\,\right]^{T}$, the columns associated with the linear velocities are removed, yielding a 2 by 6 matrix, with $\mathbf{K}$ being 2 by 4 and $\mathbf{K_i}$ being 2 by 2. The implementation of the resultant pitch and yaw controller, from the architecture in Fig. \ref{arch}, is detailed in Fig. \ref{controlsys}. 
The integral component acts on the tracking error for the pitch and yaw angles, which are a subset of $\mathbf{x_c}$ represented by $\boldsymbol\lambda = \left[\,\theta\:\:\psi\right]^T$. The attitude command $\boldsymbol\lambda_d$ is given in absolute pitch and yaw values, meaning that it has to be transformed into a relative command with respect to the values for the reference trajectory according to $\delta\boldsymbol{\lambda_d} = \boldsymbol{\lambda_d} -\mathbf{\lambda_0}$.

Once again, it is important to recall that the gain matrices $\mathbf{K}$ and $\mathbf{K_i}$ are obtained for each operating point and are scheduled throughout the flight via linear interpolation with respect to altitude.
\subsubsection{Selection of the $\mathbf{Q}$ and $\mathbf{R}$ matrices}
The design degree of freedom is the selection of the tuning matrices $\mathbf{Q}$ and $\mathbf{R}$. First of all, setting all non-diagonal entries to zero, and only focusing on the diagonal ones, allows for a more intuitive matrix selection given by the ``penalty" method \cite{friedland}. According to this method, the diagonal entries of the $\mathbf{Q}$ matrix will determine the relative importance of the state variables in terms of origin tracking performance, while the diagonal entries of the $\mathbf{R}$  matrix allow to directly adjust the control effort for each
input. Therefore, the weighting matrices have the following generic format,
\begin{subequations}
\begin{equation*}
      \mathbf{Q} = diag\,\left(\,0,\,0,\,0,\,q_q,\,q_r,\,\,q_{\theta},\,q_{\psi},\,q_{\theta_i},\,q_{\psi_i}\,\right)\,, 
\end{equation*}
\begin{equation*}
\mathbf{R} = diag\,\left(\,r_{\mu_p},\,r_{\mu_y}\,\right)\,,
\end{equation*}
\end{subequations}
where the terms associated with the linear velocities in the $\mathbf{Q}$ matrix are set to zero since those variables are not used for feedback control. The matrix entries can be iteratively adjusted by analysing the closed-loop poles and the step response of the system in the linear domain.
\subsection{Decoupled control and spin correction}
Looking at the state-space representation of the system (\ref{ss}), it is possible to identify the conditions under which it can be separated into two decoupled modes, the lateral, composed by the state vector $x_{\text{lat}} = \left[\,v\:\:r\:\:\psi\,\right]^T$ and the input $\mu_y$, and the longitudinal, composed by the state vector $x_{\text{lon}} = \left[\,u\:\:w\:\:q\:\:\theta\,\right]^T$ and the input $\mu_p$. Besides the assumption of a null roll rate, $p=0$, a condition that allows for decoupling is to consider a reference trajectory restricted to one plane, for instance, the pitch plane. By doing so, the nominal values of the lateral states and input are zero and a decoupled state-space representation is easily derived from (\ref{ss}).

Using the decoupled state-space representation, the gains for the longitudinal and lateral modes can be obtained by applying the previously detailed LQI control law, particularized to the state-space matrices associated with each individual mode. The implementation of the resultant control system is equivalent to the one in Fig. \ref{controlsys}, but now each control input is calculated separately using the gains, state estimates, and references for each mode, yielding two decoupled scheduled controllers.
 
\subsubsection{Spin correction}
The derived control system relies on the assumption that the roll rate, or spinning motion, is null ($p=0$). This can be valid if an additional roll control system is used, for instance through reaction control devices. However, it can not be guaranteed that spinning motion does not occur, given that such system can be designed to limit and not eliminate spin, or that disturbances may cause its appearance. Furthermore, it is a possibility to only have pitch and yaw control and use the spinning motion for passive stabilization through the gyroscopic effect. In this way, it is important to consider the possibility of a non-zero roll rate, $p$, and add the necessary corrections to the system so that it can still perform under that condition. In this work, we decided to correct the actuation given by the original control law, and not to rewrite the linearized dynamics including the roll rate $p$ and derive a new control law.

\begin{figure}[h]
    \centering
    \includegraphics{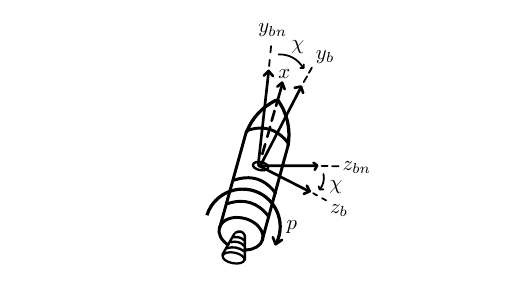}
    \caption{Non-spinning frame \{BN\}}
    \label{spincorr}
    \end{figure}

\begin{figure}[h]
    \centering
    \includegraphics{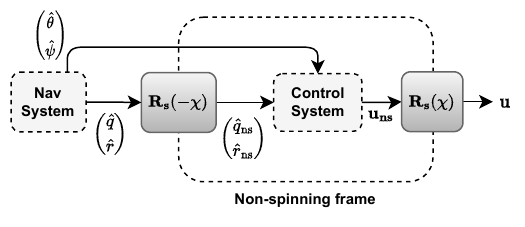}
    \caption{Spin correction for the control system.}
    \label{spincorrscheme}
    \end{figure}

Firstly, an additional frame of reference is defined: the non-spinning frame \{BN\}. This frame of reference is attached to the body but it does not rotate with respect to the x-axis, which is the spinning axis of rotation in the original body frame \{B\}. This is the frame where the states used in feedback, $\mathbf{x_c}$, and the control inputs, $\mathbf{u}$, will be defined according to the original control law. With the appearance of spinning motion, the body frame will rotate with respect to the non-spinning frame, with the instant angle of rotation being represented by $\chi$, as depicted in Fig. \ref{spincorr}. In most scenarios, $\chi$ will be very similar to $\phi$, since for small angles the roll angle approximately coincides with the x-axis body rotation.
The appearance of the angle $\chi$ means that the TVC actuation is rotated, as well as the measurements of the pitch and yaw angular velocities, both expressed in the body frame. Therefore, the estimates of the angular velocities, $\hat{q}$ and $\hat{r}$, have to be translated from \{B\} to \{BN\} before passing to the control system, and that the input vector computed in the non-spinning frame, $\mathbf{u_{ns}}$, has to be translated to \{B\}, according to scheme in Fig. \ref{spincorrscheme}.
These translations are simply given by a positive or negative instantaneous rotation of $\chi$ around the x-axis:
\begin{equation*}
\mathbf{R_s}(\chi) = \begin{bmatrix}
c_\chi & -s_\chi\\
s_\chi & c_\chi
\end{bmatrix}\,,\hspace{10pt}
\mathbf{R_s}(-\chi) = \mathbf{R_s}^T(\chi)\,,
\end{equation*}
yielding,
\begin{subequations}
\begin{equation*}
\begin{pmatrix}
\hat{q}_{ns}\\
\hat{r}_{ns}
\end{pmatrix}=
\begin{pmatrix}
\hat{q}\,c_\chi+\hat{r}\,s_\chi\\
-\hat{q}\,s_\chi+\hat{r}\,c_\chi
\end{pmatrix},
\end{equation*}
\begin{equation*}
\begin{pmatrix}
\mu_p\\
\mu_y
\end{pmatrix}=
\begin{pmatrix}
\mu_{p_{ns}}\,c_\chi-\mu_{y_{ns}}\,s_\chi\\
\mu_{p_{ns}}\,s_\chi+\mu_{y_{ns}}\,c_\chi
\end{pmatrix}.
\end{equation*}
\end{subequations}
Since the Euler angles are given in the inertial frame, no correction is needed.

With this correction method, both the coupled dynamics caused by the spinning motion and the potential lack of axial symmetry are disregarded by the control system. Therefore, its validity has to be verified for the vehicle under study, taking into account the maximum expected spin rate. 

\section{Implementation in simulation}
\label{sec:implement}
To test the proposed architecture, a simulation model transcribing the complete, non-linear, derived physical one was implemented in Matlab\&Simulink\textsuperscript{\circledR{}}. Additionally, a reference vehicle had to be selected, as well as a reference trajectory. In this Section, the reference vehicle, trajectory, and the chosen architecture parameters are detailed.   

\subsection{Reference vehicle}
The reference vehicle was obtained through a preliminary design of a low-cost, solid motor rocket to serve as a testing platform for TVC technology. The vehicle is designed to have a burning phase coinciding with
the full duration of the climb, so that TVC can be used to control its attitude
up to apogee. It is also required that the terminal velocity is inside a safe range to allow the correct activation of the recovery system. To meet these design requirements, the thrust produced by the motor is adjusted by iteratively testing different solid motor parameters, and the flight for a vertical undisturbed trajectory is simulated resorting to the simulation model.  Tables
\ref{rocket_char} and \ref{vert_traj_sim} respectively present the main vehicle characteristics and the simulation
results.
\begin{table}[ht]
    \begin{minipage}{.4\linewidth}
        \begin{tabular}{lc}
   			\toprule
   			 Total mass           & 82.9\,kg \\
    			 Dry mass             & 40.0\,kg   \\
    		     Length               & 3.57\,m  \\
    		     Max diameter         & 24\,cm  \\
    		     \bottomrule
        \end{tabular}
           \caption{Main vehicle characteristics}
   \label{rocket_char}
    \end{minipage}%
    \hspace{14pt}
    \begin{minipage}{.55\linewidth}
     \begin{tabular}{lccc}
    \toprule
    Apogee           &  4945\,m \\
    Max velocity           & 82\,m/s \\
    Max acceleration  &  1.7\,m/s$^2$\\
    Time to apogee & 100\,s \\
    Velocity at burnout & 27\,m/s\\
    \bottomrule
  \end{tabular}
          \caption{Vertical trajectory parameters}     
       \label{vert_traj_sim}
    \end{minipage} 
\end{table}

\subsection{Reference trajectory}
\begin{figure}[h]
    \centering
    \includegraphics{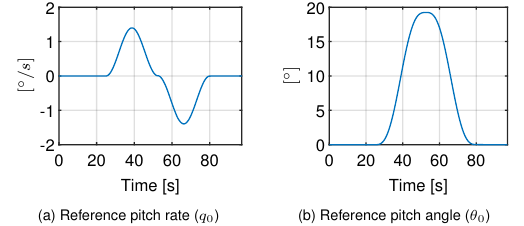}
    \caption{Reference pitch rate ($q_0$) and angle ($\theta$) over time.}
    \label{nompitch}
    \end{figure}

\begin{figure}[h]
    \centering
    \includegraphics{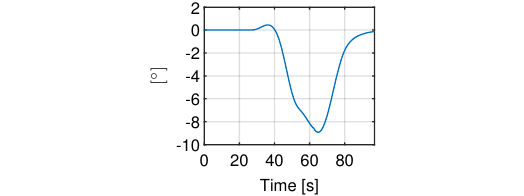}
    \caption{Nominal pitch control input ($\mu_{p_0}$)}
    \label{nompitchinput}
    \end{figure}

Regarding the attitude reference that defines the reference trajectory, a varying pitch trajectory, in which the controller restricts the motion to the pitch plane (yaw equal to zero) and makes the vehicle deviate from the vertical to later recover it, is selected. In this way, it is ensured that the apogee is reached further away from the launch site, increasing safety. Figure \ref{nompitch} shows the reference pitch rate and angle over time. 
\\
The feedforward control inputs are computed as stated in subsection \ref{feedforward}, yielding the nominal actuation present in Fig. \ref{nompitchinput}. The PID gains were set to $k_p = -10$, $k_i=-20$, and $k_d=-5$.

\subsection{Architecture parameters}
In this subsection, the architecture parameters, obtained after tuning, are presented. These include the gains for both the navigation and control systems, as well as the model used to represent the actuator's dynamics.
\subsubsection{Navigation system parameters}
After tuning both complementary filters through their respective $\mathbf{Q}$ and $\mathbf{R}$ covariance matrices, the following constant gain matrices were obtained:
\begin{equation*}
\mathbf{L}_{\text{acf}} = \begin{bmatrix}
0.5\,\mathbf{I}_{3}\\
-\mathbf{I}_{3}
\end{bmatrix}\,,\hspace{5pt}
\mathbf{L_1} = \mathbf{L_2} = \mathbf{I}_{3}\,,\hspace{5pt}
\mathbf{L_3} = -0.5\,\mathbf{I}_{3}  
\end{equation*}

\subsubsection{Control system parameters}
\begin{figure}[h]
    \centering
    \includegraphics{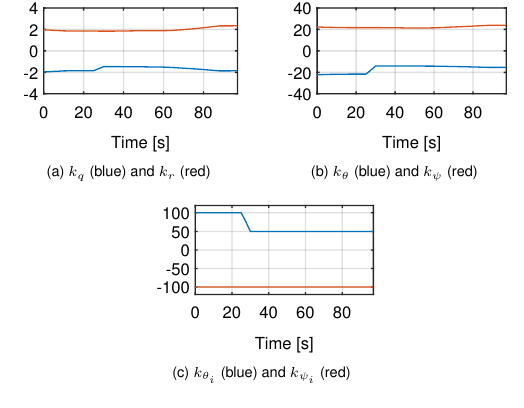}
    \caption{Controller gains over time.}
    \label{gainsfig}
    \end{figure}

For the control system, the decoupled version was used, yielding two separate gain matrices, one for for each mode: $\mathbf{K_{\text{lon}}} = \left[\,k_q\:\: k_\theta\:\: k_{\theta_i}\,\right]^{T}$ and $\mathbf{K_{\text{lat}}} = \left[\,k_r\:\: k_\psi\:\: k_{\psi_i}\,\right]^{T}$. Figure \ref{gainsfig} details the evolution of the gains throughout the nominal trajectory, obtained after tuning the $\mathbf{Q}$ and $\mathbf{R}$ matrices. 
The gains remain approximately constant given that the tuning matrices were left constant for all operating points, except for the ones associated with the longitudinal mode during the varying pitch section, which were tuned in order to reduce the control effort and avoid saturation. 

The actuator dynamics are modelled using a continuous time first-order transfer function for each input ($\mu_p$ and $\mu_y$), considering a servo-actuated system. The transfer function is given by
\begin{equation*}
 \mu_r = \frac{1}{\tau\,s + 1}\,\mu\,
\end{equation*}
where $\mu_r$ is the actuator angular response and $\tau$ is the time constant. Additionally, servo motors normally have a saturation value for the rotation velocity, which can be modelled by a rate limiter block in Simulink. The time constant and angular velocity limit values were retrieved from typical high grade servo motors, and are equal to $0.02\,$s and 1 full rotation per second, respectively.   

\section{Linear domain analysis}
\label{sec:linanalysis}
Using the linear representation of the system (\ref{ss}) and the reference values its states, inputs, and parameters, it is possible to derive both the open-loop and closed-loop stability and response in the linear domain.

For a time-varying system, determining the location of the poles throughout the reference trajectory does not provide a mathematical stability proof, however, the study is carried out to understand the behaviour of the system throughout the flight. Given the symmetry of the vehicle, and the fact that the motion is restricted to the pitch plane, the study is performed for the longitudinal mode. 
\subsection{Open-loop stability}

\begin{figure}[h]
    \centering
    \includegraphics{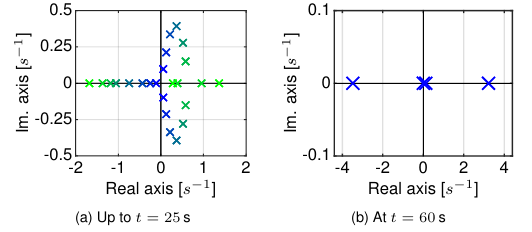}
    \caption{Open-loop poles.}
    \label{poles}
    \end{figure}

Figure \ref{poles} details the pole evolution (from blue to green) during the initial vertical section (up to $25\,$s) and the poles at $t = 60\,$s, which serves as example for the distribution type during the varying pitch section.
By evaluating the location of the open-loop poles some conclusions can be made. Firstly, the system is naturally unstable, which was expected due to negative static stability margin caused by the absence
of aerodynamic fins. Secondly, the system displays natural oscillatory behaviour during the first seconds, after which all poles are located in the real axis. Finally, it is concluded that the velocity of the vehicle is a driving factor for the response of the system: at higher velocities the system is seen to have higher magnitude poles and hence faster dynamics.

\subsection{Closed loop stability and response}

\begin{figure}[h]
    \centering
    \includegraphics{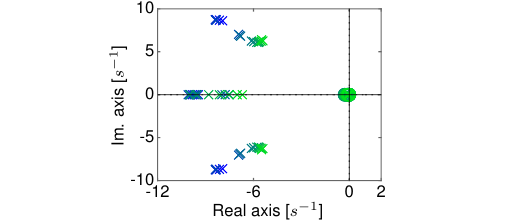}
    \caption{Closed-loop poles and zeros.}
    \label{polesclosed}
    \end{figure}

By closing the loop with the derived control law, the closed-loop poles and zeros can be determined for the different operating points of the reference trajectory. Figure \ref{polesclosed} displays the poles and zeros for the longitudinal mode, for all the selected operating points.
The control law allowed to stabilize all operating points, placing all closed-loop poles in the left-hand side of the complex plane. The pole-zero cancellation of the poles and zeros approximately located at the origin is noted. For each operating point, the relevant poles correspond to a pair of conjugated complex poles and a real pole, all in the left-hand side of the complex plane. The complex poles are expected to cause oscillatory behaviour in the response of the system, nonetheless, it was the ideal compromise found, during the design iteration, between limiting oscillations while keeping a fast settling time. 

\begin{figure}[h]
    \centering
    \includegraphics{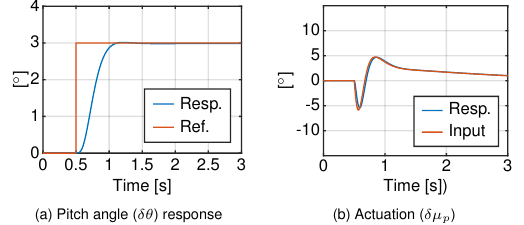}
    \caption{Response to a $3\,^\circ$ step in pitch angle.}
    \label{stepfig}
    \end{figure}

The step response was also analysed. Figure \ref{stepfig} displays the response to a step request of $3$ degrees in pitch angle, and the associated control input variation, at $t = 60\,$s, as exemplification of the performance.
Table \ref{steplintab} details some key parameters of the closed-loop system step response in the linear domain for distinct operating points. 
\begin{table}[ht]
\begin{center}
\begin{minipage}{252pt}
 \begin{tabular}{@{}lccc}
    \toprule
      Op. point   & Rise time (s) & Settling time (s) & Overshoot (\%) \\
    \midrule
      $t = 5$\,s &  0.2686    &    0.4461  &   0.5710  \\
      $t = 35$\,s  &   0.3401     & 0.5723   &  0.1239 \\
      $t = 65$\,s & 0.3278 & 0.5303 & 1.7586\\
      $t = 95$\,s & 0.3667 & 0.6052 & 0.7995 \\
    \bottomrule
  \end{tabular}
  \caption{Closed-loop step response parameters}\label{tab:clrespparam}
  \label{steplintab}
\end{minipage}
\end{center}
\end{table}
The response is seen to be approximately constant for all operational regimes, being fast and having limited overshoot.

\section{Simulation results}
\label{sec:simres}

\begin{figure}[h]
    \centering
    \includegraphics{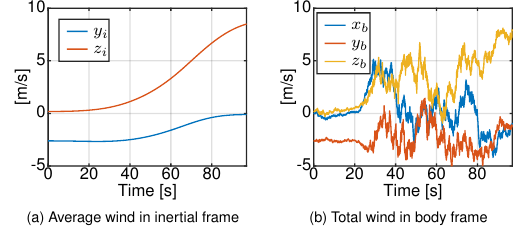}
    \caption{Average horizontal wind and total wind.}
    \label{windfig}
    \end{figure}

In this Section, the results obtained using the proposed architecture in the simulation environment are presented. As feedforward control is sufficient to stabilize the plant for a disturbance-free flight, stochastic wind was added to test feedback control. Wind was simulated by using the horizontal average wind model with gusts added from the Dryden model, both available as Simulink blocks. Figure \ref{windfig} displays the average horizontal wind and the total wind (gusts included), used in simulation.
\subsection{Navigation system}
In the following two subsections, the estimation results, obtained for the reference trajectory, are shown for each complementary filter. 
\subsubsection{ACF}
\begin{figure}[h]
    \centering
    \includegraphics{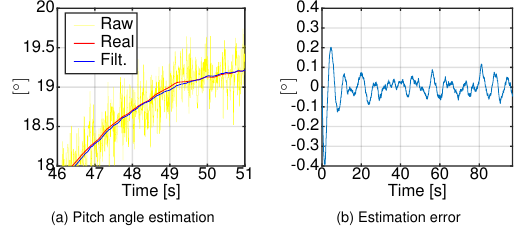}
    \caption{ACF pitch angle estimation.}
    \label{acfpitch}
    \end{figure}

The ACF is able to remove the noise from the Euler angles readings and to correct the bias of the gyroscope, providing an accurate estimate on the attitude of the vehicle. Figure \ref{acfpitch} (a) displays the estimation results for the pitch angle in a zoomed interval, while Fig. \ref{acfpitch} (b) presents the pitch estimation error for the entire flight.
It is possible to infer that an accurate estimate is obtained in the presence of a noisy measurement, and that the estimation error, after initial stabilization, is limited to $\pm 0.1$ degrees.   

\begin{figure}[h]
    \centering
    \includegraphics{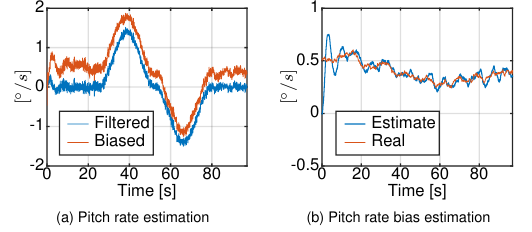}
    \caption{ACF pitch rate estimation.}
    \label{acfrate}
    \end{figure}

Figure \ref{acfrate} illustrates the pitch rate estimation. In Fig. \ref{acfrate} (a) it is seen that the bias is removed from the gyroscope measurement, and in Fig. \ref{acfrate} (b) the respective bias estimation is shown.
    
\subsubsection{PCF}
\begin{figure}[h]
    \centering
    \includegraphics{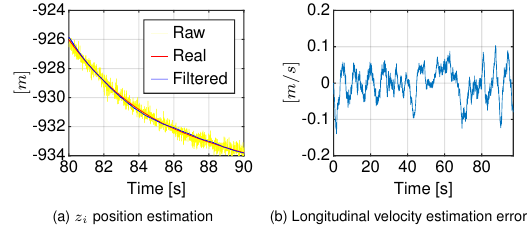}
    \caption{PCF estimation.}
    \label{pcfres}
    \end{figure}

Regarding the PCF, it was also possible to verify its correct functioning by analysing the position and velocity estimates. Figure \ref{pcfres} (a) shows a zoomed section of the crossrange position ($z_i$) estimation to better understand the filtering done by the PCF, and Figure \ref{pcfres} (b) details the longitudinal velocity ($u$) estimation error for the entire flight.  
The position filtering is able to reject the noise from the measurements while maintaining good accuracy with respect to the true value. The filter also provides accurate estimates on velocity.
\subsection{Control system}

\begin{figure}[h]
    \centering
    \includegraphics{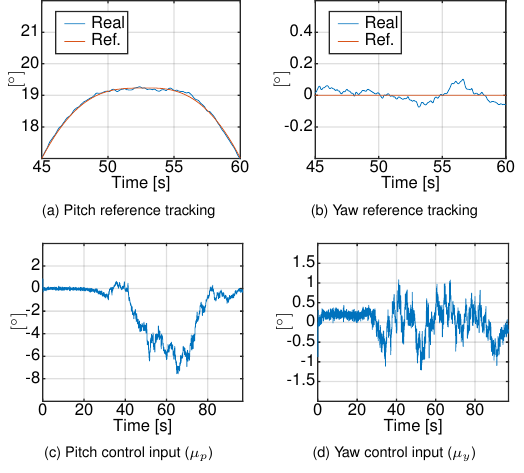}
    \caption{Attitude reference tracking.}
    \label{controlfigs}
    \end{figure}

With the implementation of the control system, the full architecture was able to reject the external wind perturbation. Figure \ref{controlfigs} presents the simulation results for the pitch and yaw angles reference tracking, for a portion of the flight, with the navigation system included in the loop. It also details the actuation by the TVC system.
Table \ref{trackingtab} presents the detailed results, in terms of the sum squared tracking errors and the root mean square of the actuation signals, when using the derived LQI control law, as well as the ones obtained when using a PID controller per degree of freedom (pitch and yaw). The results are shown with and without including measurement noise and the estimator in the loop. 
\begin{table}[ht]
\begin{center}

\begin{tabular}{@{}lccccc@{}}
    \toprule
         & \multicolumn{2}{c}{Exact state} && \multicolumn{2}{c}{Estimated state} \\
         \cmidrule{2-3}
         \cmidrule{5-6}
         & LQI & PID && LQI & PID\\
    \midrule
     $\Sigma {\theta_e}^2 $ &  1.83   & 5.46  && 12.69 & 16.86\\
      $\Sigma {\psi_e}^2 $  &  0.33   &  2.67 && 11.12 & 13.60 \\
      $\delta_{\mu_1,rms}$   & 1.40 & 1.40 &&  1.40 & 1.56\\
       $\delta_{\mu_2,rms}$ &  0.37   & 0.38   &&  0.38 & 0.77\\
       \bottomrule
  \end{tabular}
  \caption{Tracking error and control effort (angles expressed in degrees).}
  \label{trackingtab}
\end{center}
\end{table}
It is noted that LQI control provides better attitude tracking for a similar control effort with respect to PID control. As expected, there is a decrease in performance when measurement noise is added, yet, it is still considered satisfactory.

\begin{figure}[h]
    \centering
    \includegraphics{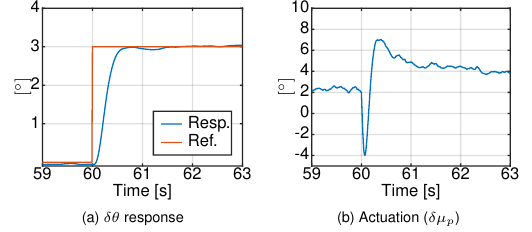}
    \caption{Response to a $3\,^{\circ}$ step in $\delta \theta$}
    \label{stepresim}
    \end{figure}

The step response is also analysed to determine if the system is able to track deviations from the reference condition. Figure \ref{stepresim} illustrates the results for the same instant and request as shown for the linear domain (Fig. \ref{stepfig}).
It is possible to verify that the step response performance is similar to the one found for the linear domain, apart from disturbance/noise induced irregularities.

\subsubsection{Robustness analysis}

Finally, a robustness analysis was performed to determine the robustness of the architecture to model uncertainties. Several system parameters, including mass, inertias, Thrust, centre of mass position, and aerodynamic coefficients, were altered independently, inside admissible ranges in terms of percentage of the original value. The system showed sufficient robustness, being able to stabilize the plant for all the variations under study. The parameter which demonstrated the highest influence in the performance of the control system was the position of the centre of mass ($x_{cm}$), with the results shown in Table \ref{cmvary}.  
\begin{table}[ht]
\begin{center}
 \begin{tabular}[]{lccccc}
    \toprule
      (x) $x_{cm}$   & 0.8 &0.9 & 1 & 1.1 & 1.2\\
    \midrule
     $\Sigma {\theta_e}^2 $ &  3.13  &  6.31 & 12.69 & 22.21 & 39.56 \\
      $\Sigma {\psi_e}^2 $  &  2.59   & 5.78  & 11.12 & 19.24 & 37.31 \\
      $\delta_{\mu_1,rms}$   & 0.79 & 1.06 & 1.40 & 1.79 & 2.28\\
       $\delta_{\mu_2,rms}$ &  0.21   & 0.30  &  0.38 & 0.48 & 0.63 \\
       \bottomrule
  \end{tabular}
  \caption{Attitude tracking performance for different $x_{cm}$}\label{cmvary}
\end{center}
\end{table}
A lower value, meaning a position closer to the tip of the rocket, causes the moment arm for the thrust vector actuation to be higher, which increases the control authority. At the same time, the natural instability of the rocket reduces. In this way, the tracking performance increases when the centre of mass moves closer to the tip, while the control effort decreases. 

\section{Conclusions}
\label{sec:conclusions}

With the conclusion of this work, it is possible to state that the primary goal has been achieved: the
successful design of an integrated TVC and state estimation architecture, applicable to low-cost small-scale launch vehicles. An original linear state-space representation was derived for the generic thrust-vector-controlled launch vehicle, which served as foundation for the architecture design. The navigation system relies on readily available components, providing accurate state estimates by removing measurement noise and bias. The use of linear Kalman filtering with pre-calculated gains reduces the required computational power on board. The control system, based on the scheduling of pre-calculated gains with an LQI control law, ensured satisfactory attitude reference tracking performance and robustness to model uncertainties. The integrated architecture was tested in simulation, using the derived non-linear model for the vehicle dynamics and kinematics, yielding satisfactory performance.  As future work, a final validation of the proposed architecture shall occur through the use of small-scale rocket prototypes, before its implementation in a real launch vehicle.

\section*{Declaration of competing interests}
The authors declare that they have no known competing financial interests or personal relationships that could have appeared to influence the work reported in this paper.

\section*{Acknowledgements}
This work is financed by national funds through FCT – Foundation for Science and Technology, I.P., through IDMEC, under LAETA, projects UIDB/50022/2020 and CAPTURE PTDC/EEI-AUT/1732/2020. 



%%       \citet{<label>} ==> Jones et al. [21]
%%       \citep{<label>} ==> [21]
%%

  \bibliographystyle{elsarticle-num-names} 
  \bibliography{bibliography}


%\begin{thebibliography}{00}



%\end{thebibliography}
\end{document}